\documentclass[aps,pra,reprint]{revtex4-2}

\usepackage{graphicx}
\usepackage{bm}
\usepackage{bbold}
\usepackage{amssymb}
\usepackage{amsmath}

\renewcommand{\vec}[1]{{\mathbf #1}}

\bibpunct{[}{]}{,}{n}{}{}

\begin{document}

\title{Topological transitions and Anderson localization of light in disordered atomic arrays}

\author{S.E. Skipetrov}
\email[]{Sergey.Skipetrov@lpmmc.cnrs.fr}
\affiliation{Univ. Grenoble Alpes, CNRS, LPMMC, 38000 Grenoble, France}

\author{P. Wulles}
\affiliation{Univ. Grenoble Alpes, CNRS, LPMMC, 38000 Grenoble, France}

\date{\today}

\begin{abstract}
We explore the interplay of disorder and topological phenomena in honeycomb lattices of atoms coupled by the electromagnetic field. On the one hand, disorder can trigger transitions between distinct topological phases and drive the lattice into the topological Anderson insulator state. On the other hand, the nontrivial topology of the photonic band structure suppresses Anderson localization of modes that disorder introduces inside the band gap of the ideal lattice. Furthermore, we discover that disorder can both open a topological pseudogap in the spectrum of an otherwise topologically trivial system and introduce spatially localized modes inside it.
\end{abstract}

\maketitle

\section{Introduction}
\label{secintro}

Topological photonics \cite{lu14,ozawa19,segev21} is a rapidly developing field of research with great prospects for development of novel light sources \cite{bahari17,harari18,zeng20}, integrated photonic circuits \cite{zhao19,ma19a}, and quantum information processing devices \cite{wang19,tschernig21,chen21}. It is largely inspired by topological physics of charged fermions (electrons) in condensed matter \cite{haldane17,wen17,hasan10} but have its own peculiarities originating from the differences between electrons and photons: photons are bosons with neither mass nor charge but with a polarization \cite{haldane08,raghu08,wang08,wang09,rechtsman13}. The key feature of topological phenomena is their robustness with respect to defects and disorder. It allows one to apply theoretical concepts and designs developed for idealized model systems to real-world materials and devices that can be manufactured and controlled only with a limited precision, especially in miniaturized setups. More recently, it has been realized that in the context of topological physics, disorder can be not only a boring nuisance but also a useful resource capable of conferring nontrivial topological properties to an otherwise topologically trivial system by triggering a quantum phase transition towards a so-called topological Anderson insulator (TAI) phase \cite{li09,groth09,liu17,stutzer18,meier18,liu20}. Despite its name, TAI does not rely on Anderson localization phenomenon and can actually be understood in the framework of an effective-medium theory \cite{groth09}. The interplay between genuine Anderson localization and topological phenomena is still largely unexplored.

Most successful experimental realizations of topological photonic phenomena involve two-dimensional (2D) arrays of microwave resonators \cite{wang09,ma19b,liu20,ma20,reisner21} or parallel waveguides \cite{rechtsman13,stutzer18}. Recently, a planar array of two-level atoms arranged in a 2D honeycomb lattice embedded in three-dimensional (3D) free space (see Fig.\ \ref{fig_honeycomb}) has been proposed as an alternative, quantum-optical platform \cite{perczel17}. Its potential advantages include the possibility of optically addressing individual atoms and exploiting the strong optical nonlinearity as well as the quantum nature of atomic transitions.
In this paper, we study the interplay of topology and disorder in this system. On the one hand, we show that disorder can trigger topological transitions between trivial and nontrivial topological phases by closing or opening topological (pseudo)gaps in the spectrum. The existence of the optical TAI phase is demonstrated. On the other hand, the type of the topological phase is shown to affect the degree of spatial localization of optical modes introduced into the gap by disorder. In TAI phase, disorder plays a two-fold role by being at the origin of both the pseudogap (that would not exist without disorder) and the spatially localized states arising inside it.
Previous works focused on disorder in on-site potential energies \cite{li09, groth09} or nearest-neighbor hopping terms \cite{meier18}  for a quantum particle on a lattice, as well as in permittivity  \cite{liu17, stutzer18} or orientation \cite{liu20} of optical scattering units. Demonstrations of topologically protected edge modes typically involved a single localized defect \cite{perczel17, liu20}. 
In contrast, we consider disorder in positions of many identical atoms which is easier to implement experimentally. To our knowledge, the impact of such a disorder on the topological properties of optical systems and its capacity of inducing TAI phase have never been demonstrated before.  In addition, our calculation accounts for the non-Hermitian character of the considered physical system in a realistic way.

\section{Light in a honeycomb atomic lattice}
\label{sechoney}

\begin{figure}
\includegraphics[width=0.9\columnwidth]{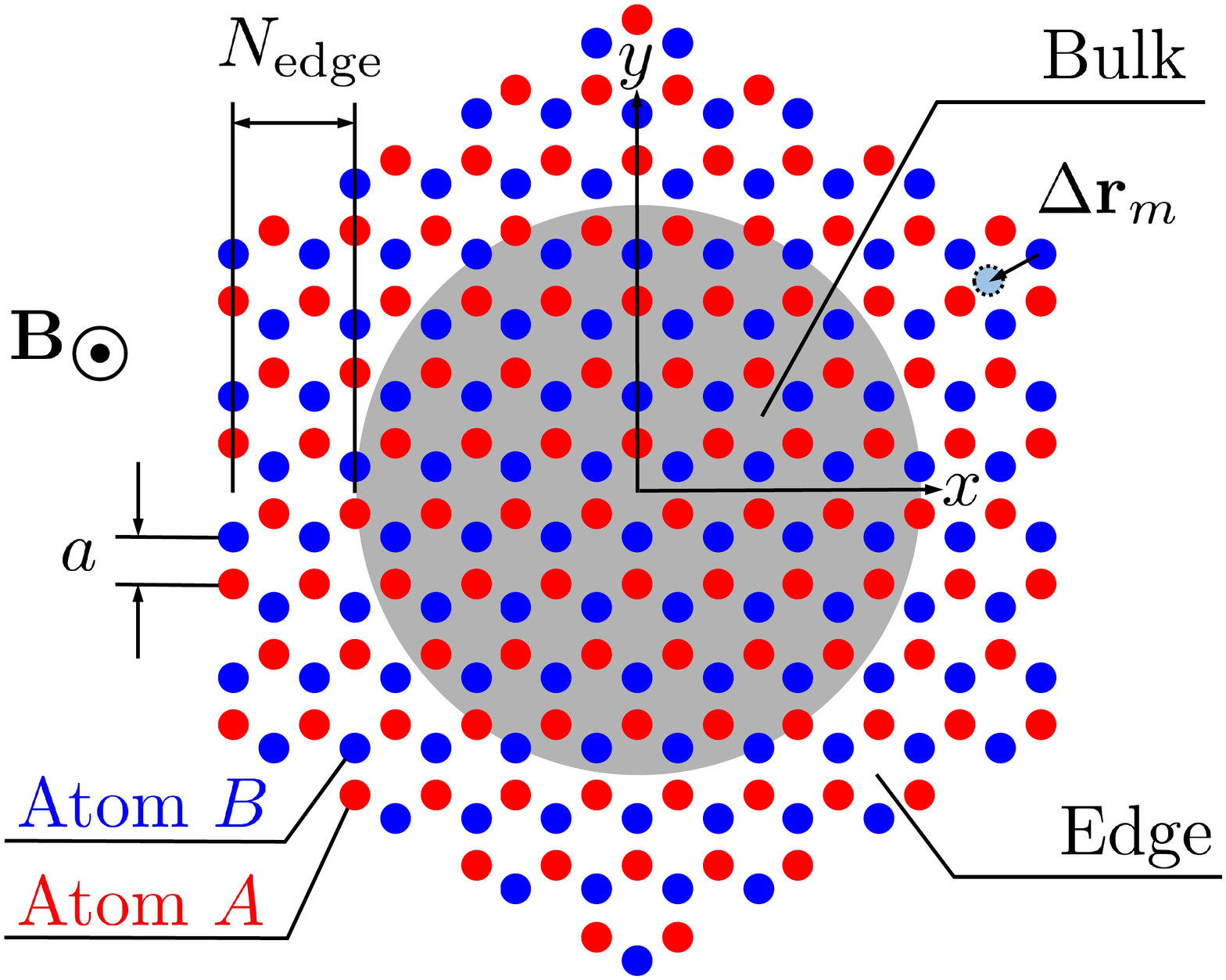}
\vspace{-2mm}
\caption{\label{fig_honeycomb}
Honeycomb lattice of two-level atoms. Atoms $A$ (red) and $B$ (blue) form a unit cell. We split the finite-size lattice in  circular bulk (shaded in grey) and remaining edge parts; the thinnest part of the edge has a width of $N_{\text{edge}}$ atomic layers. Time-reversal symmetry can be broken by a magnetic field $\vec{B}$ perpendicular to the plane $xy$ of the lattice. Inversion symmetry can be broken by taking atoms $A$ and $B$ different. Disorder is introduced by displacing atoms by random distances $\Delta r_m \in [0, Wa]$ in random directions.}
\end{figure}

We consider a honeycomb lattice of $N \gg 1$ immobile atoms located in the $xy$ plane at positions denoted by $\vec{r}_m = \{x_m, y_m \}$, $m = 1, \ldots, N$.
The unit cell of the honeycomb lattice is composed of a pair of atoms belonging to two triangular sublattices $A$ and $B$ separated by the lattice parameter $a$ and shown by different colors in Fig.\ \ref{fig_honeycomb}. We assume that the ground and excited states of the atoms have total angular momenta $J_g = 0$ and $J_e = 1$, respectively, so that the excited state is triply degenerate. Resonance frequencies of atoms $A$ and $B$ are $\omega_A$ and $\omega_B$, and the atomic lattice is placed in a uniform external magnetic field $\vec{B}$ directed along the axis $z$ perpendicular to the plane $xy$ of the lattice. The atoms interact via the electromagnetic field and the excitations polarized in the $xy$ plane ($\sigma = \pm 1$, TE modes) are decoupled from those polarized along the $z$ axis ($\sigma = 0$, TM modes). Here we consider only TE modes; TM modes can be studied separately but their physics is less interesting because they are not affected by the magnetic field. With all these assumptions, the effective Hamiltonian of the lattice takes the form of a $2N \times 2N$ 
non-Hermitian matrix $\hat G$ composed of $2 \times 2$ blocks ${\hat G}_{mn}$ describing the interaction of atoms $m$ and $n$ \cite{skip15,perczel17}:
\begin{eqnarray}
{\hat G}_{mn} &=& \delta_{mn} \left[\left( i \pm 2 \Delta_{AB} \right) \mathbb{1} + 2 {\hat \sigma}_z \Delta_{\vec{B}} \right]
\nonumber \\
&-& \frac{6\pi}{k_0}(1 - \delta_{mn}) {\hat d}_{eg} \hat{\cal G}(\vec{r}_m - \vec{r}_n) {\hat d}_{eg}^{\dagger} \;\;\;\;
\label{greene}
\end{eqnarray}
where $\Delta_{AB} = (\omega_B - \omega_A) /2\Gamma_0$ is the frequency detuning between $A$ and $B$ sublattices in units of the decay rate of the excited states $\Gamma_0$ (assumed identical for atoms $A$ and $B$), ``$+$'' or ``$-$'' signs correspond to atoms $A$ and $B$, respectively, $\Delta_{\vec{B}} = \mu_B B/\Gamma_0$ is the Zeeman shift due to the magnetic field (again, in units of $\Gamma_0$), $\mu_B$ is the Bohr magneton, $\mathbb{1}$ is the $2 \times 2$ identity matrix and ${\hat \sigma}_z$ is the third Pauli matrix. ${\hat{\cal G}}(\vec{r})$ is the dyadic Green's function of Maxwell equations:
\begin{eqnarray}
\hat{{\cal G}}(\vec{r}) &=& -\frac{e^{i k_0 r}}{4 \pi r}
\left[ P(i k_0 r) \mathbb{1}
+ Q(ik_0 r) \frac{\vec{r} \otimes \vec{r}}{r^2} \right]
\label{greena}
\end{eqnarray}
describing the propagation of electromagnetic waves between atoms in the basis of linear polarizations.
Here $k_0 = \omega_0/c = 2\pi/\lambda_0$, $\omega_0 = (\omega_A + \omega_B)/2$,  $P(x) = 1 - 1/x + 1/x^2$, and $Q(x) = -1 + 3/x - 3/x^2$.
\begin{eqnarray}
{\hat d}_{eg} = \frac{1}{\sqrt{2}}
\begin{bmatrix}
1 & i \\
-1 & i \\
\end{bmatrix}
\label{dmatrix}
\end{eqnarray}
transforms $\hat{\cal G}(\vec{r}_m - \vec{r}_n)$ into the basis of circular polarizations $\sigma = \pm 1$.

Topological properties of an ideal honeycomb atomic lattice are governed by the balance between time-reversal and inversion symmetry breaking controlled by the parameters $\Delta_{\mathbf{B}}$ and $\Delta_{AB}$, respectively. This is very similar to Haldane's model \cite{haldane88} despite the long-range coupling between atoms in our system. The case $\Delta_{AB} = 0$ (identical atoms $A$ and $B$) has been studied in Ref.\ \cite{perczel17}: the magnetic field opens a topological band gap around $(\omega-\omega_0)/\Gamma_0 \approx 7$; the gap width $\Delta_{\mathrm{Gap}} = 2|\Delta_{\vec{B}}|$
(in units of $\Gamma_0$)
is bounded by a maximum value that scales as $(a/\lambda_0)^{-3}$.
This scaling suggests that opening of a band gap for light in a honeycomb atomic lattice is due to dipole-dipole interactions between neigboring atoms. Typically, for $a = \lambda_0/20$ considered below, the maximum width of the gap is $\max(\Delta_{\mathrm{Gap}}) \simeq 24$. However, $\max(\Delta_{\mathrm{Gap}})$ decreases to 0.1 already for $a = \lambda_0/5$ \cite{perczel17}. Extending the analysis of Ref.\ \cite{perczel17} to $\Delta_{AB} \ne 0$, we find $\Delta_{\mathrm{Gap}} = 2 ||\Delta_{\vec{B}}| - |\Delta_{AB}||$.
Calculation of Chern numbers $C$ \cite{bernevig13,fukui05} of bands shows that the gap is topological when $|\Delta_{\vec{B}}| > |\Delta_{AB}|$ and trivial otherwise.

\section{Disorder and Bott index}
\label{secbott}

Let us now turn to the main subject of this work: the impact of disorder on the topological properties. We introduce disorder by displacing atoms randomly from their positions in the honeycomb lattice and study the spectrum  $\omega_{\alpha} = \omega_0 - (\Gamma_0/2)\text{Re} \Lambda_{\alpha}$ and right and left eigenvectors $|R_{\alpha} \rangle$ and $|L_{\alpha} \rangle$ of the matrix $\hat{G}$, to which we will also refer as ``quasimodes''. These are obtained by solving 
\begin{eqnarray}
{\hat G} | R_{\alpha} \rangle  &=& \Lambda_{\alpha} | R_{\alpha} \rangle
\label{eigenr}
\\
\langle L_{\alpha} | {\hat G} &=& \langle L_{\alpha} | {\tilde \Lambda}_{\alpha}^*
\label{eigenl}
\end{eqnarray}
in samples having a shape of a hexagon with arm-chair edges shown in Fig.\ \ref{fig_honeycomb}.
In honeycomb lattices with nearest-neighbor coupling, arm-chair edges are known to host no edge modes in the presence of both the time-reversal and inversion symmetries (i.e., when $\Delta_{\vec{B}} = \Delta_{AB} = 0$) \cite{nakada96, kohmoto07, bellec13}. We verified that this property is preserved in our lattice of atoms coupled by propagating electromagnetic waves. In finite-size samples with arm-chair edges, edge modes arising due to the breakdown of symmetries are easier to identify and cannot be confused with ``trivial''edge modes that may exist in samples with other types of regular (zigzag or bearded) or irregular edges.

Topological properties of the lattice are characterized by Bott index $C_B$ calculated for square samples of side $L$  \cite{loring10}:
\begin{eqnarray}
C_B(\omega) = \frac{1}{2\pi} \mathrm{Im} \mathrm{Tr} \ln [{\hat V}_X(\omega) {\hat V}_Y(\omega) {\hat V}_X^{\dagger}(\omega) {\hat V}_Y^{\dagger}(\omega)]
\label{bott}
\end{eqnarray}
where
\begin{eqnarray}
{\hat V}_{X,Y}(\omega)  &=& {\hat P}(\omega) {\hat U}_{X,Y} {\hat P}(\omega)
\label{vxy}
\\
{\hat U}_X &=& \exp(i 2\pi {\hat X}/L)
\label{ux}
\\
{\hat U}_Y &=& \exp(i 2\pi {\hat Y}/L)
\label{uy}
\end{eqnarray}
and ${\hat X}$ and ${\hat Y}$ are diagonal $N \times N$ matrices containing coordinates $x_m$ and $y_m$ of atoms,
\begin{eqnarray}
{\hat P}(\omega) = \sum_{\omega_{\alpha} \leq \omega} | R_{\alpha} \rangle \langle L_{\alpha} |
\label{proj}
\end{eqnarray}
is a projector operator on quasimodes corresponding to frequencies $\omega_{\alpha}$ below $\omega$.
The openness of the considered physical system and the resulting non-Hermitian character of the matrix $\hat{G}$ modify the definition of the projector operator (\ref{proj})  that now include both right and left eigenvectorts \cite{moiseyev11}; otherwise the definition of $C_B$ follows the original proposal \cite{loring10}. We provide details of calculations in Appendix \ref{appbott}.

Bott index $C_B \ne 0$ signals a topologically nontrivial band gap accompanied by topologically protected edge modes in a sample of finite size. For the ideal honeycomb lattice (no disorder), we verified that $C_B$ coincides with Chern number $C$ for frequencies inside the band gap. In contrast to $C$, $C_B$ has the advantage of being well-defined in the presence of disorder as well.

To illustrate our main conclusions, we present below results obtained for lattices of sizes $N = 4326$ (hexagonal samples) and 2244 (square samples) that we find to be large enough to represent the large-$N$ limit. We fix the nearest-neighbor spacing $a = \lambda_0/20$ \footnote{The reported results remain qualitatively the same for $a$ up to at least $\lambda_0/5$ beyond which the narrowness of the band gap complicates numerical analysis.}. Because topological phenomena confer special properties to edge modes, it turns out useful to analyze modes localized at the edges separately from those in the bulk of the sample. To this end, we separate the sample into ``bulk'' and ``edge'' parts by requiring bulk modes $| R_{\alpha} \rangle$ to have at least 50\% of their weight $\langle R_{\alpha} | R_{\alpha} \rangle$ inside the bulk (see Fig.\ \ref{fig_honeycomb}). On the contrary, edge modes should have more than 50\% of their weight in the edge part of the sample. The bulk part is circular and the edge has a minimum width of $N_{\text{edge}} = 4$ atomic layers (see Fig.\ \ref{fig_honeycomb}) \footnote{This choice is somewhat arbitrary but we checked that our conclusions remain qualitatively the same for $N_{\text{edge}} = 1$--5.}. Finally, ensemble averaging is performed over 50--200 independent realizations of disorder, except when indicated otherwise.

\begin{figure*}
\includegraphics[width=0.8\textwidth]{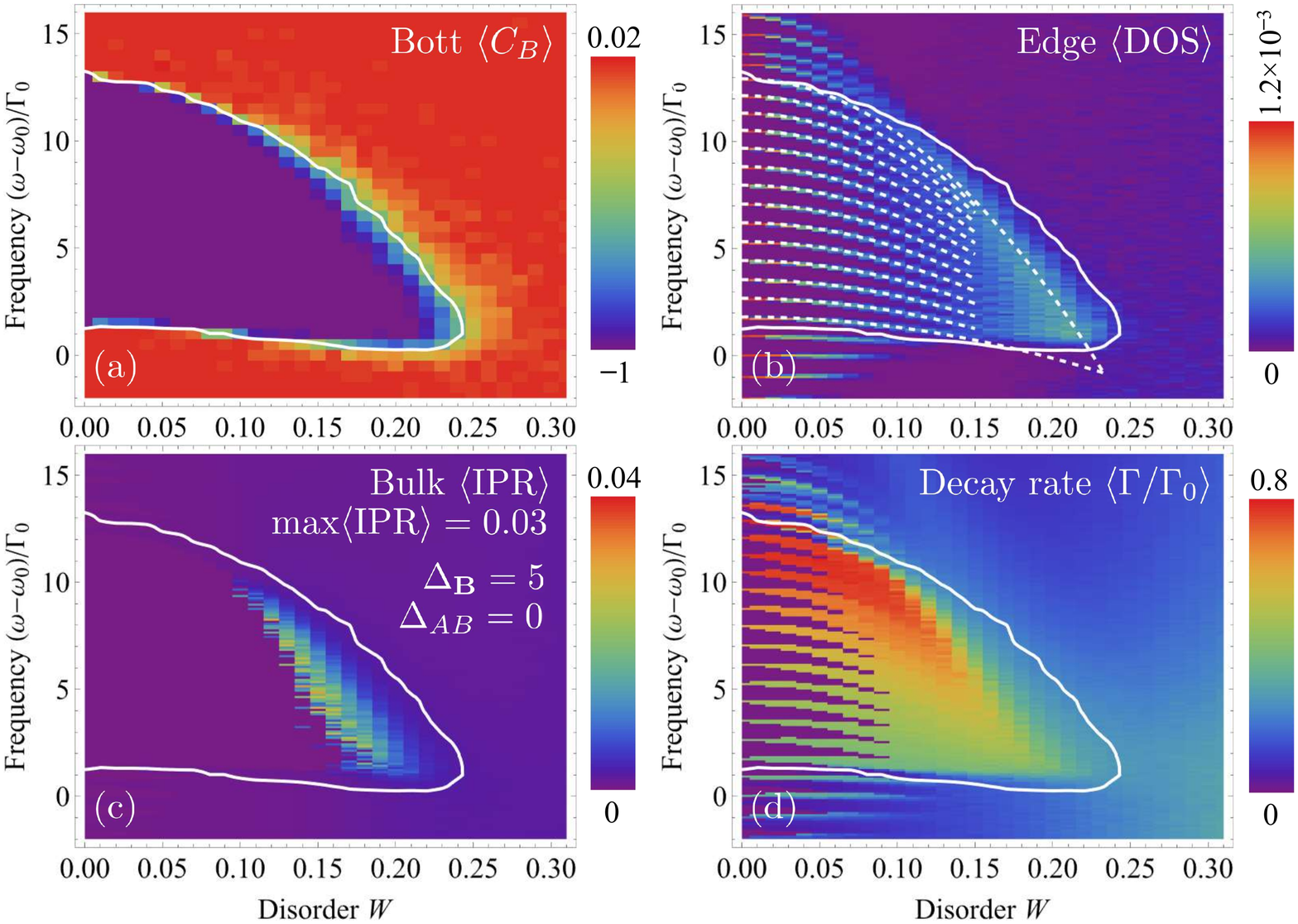}
\caption{\label{fig50}
Closing of a topological band gap by disorder. (a) Average Bott index $\langle C_B \rangle$ as a function of disorder strength $W$ for $\Delta_{\vec{B}} = 5$ and $\Delta_{AB} = 0$. The white contour reproduced for reference in all plots (a)--(c) shows the level $\langle C_B \rangle = -0.5$. (b) Average edge DOS. Dashed lines show edge modes predicted by the perturbation theory. Frequencies of the lowest- and highest-frequency edge modes traced up to higher values of $W$ approximate band edges. (c) Average bulk IPR. Deep violet color corresponds to either very low IPR or $\text{DOS} = 0$ (no modes). (d) Average quasimode decay rate in units of the natural decay rate $\Gamma_0$ of an isolated atom.
}
\end{figure*}

\section{Closing of a topological band gap by disorder}
\label{secclosing}

\begin{figure*}
\includegraphics[width=0.8\textwidth]{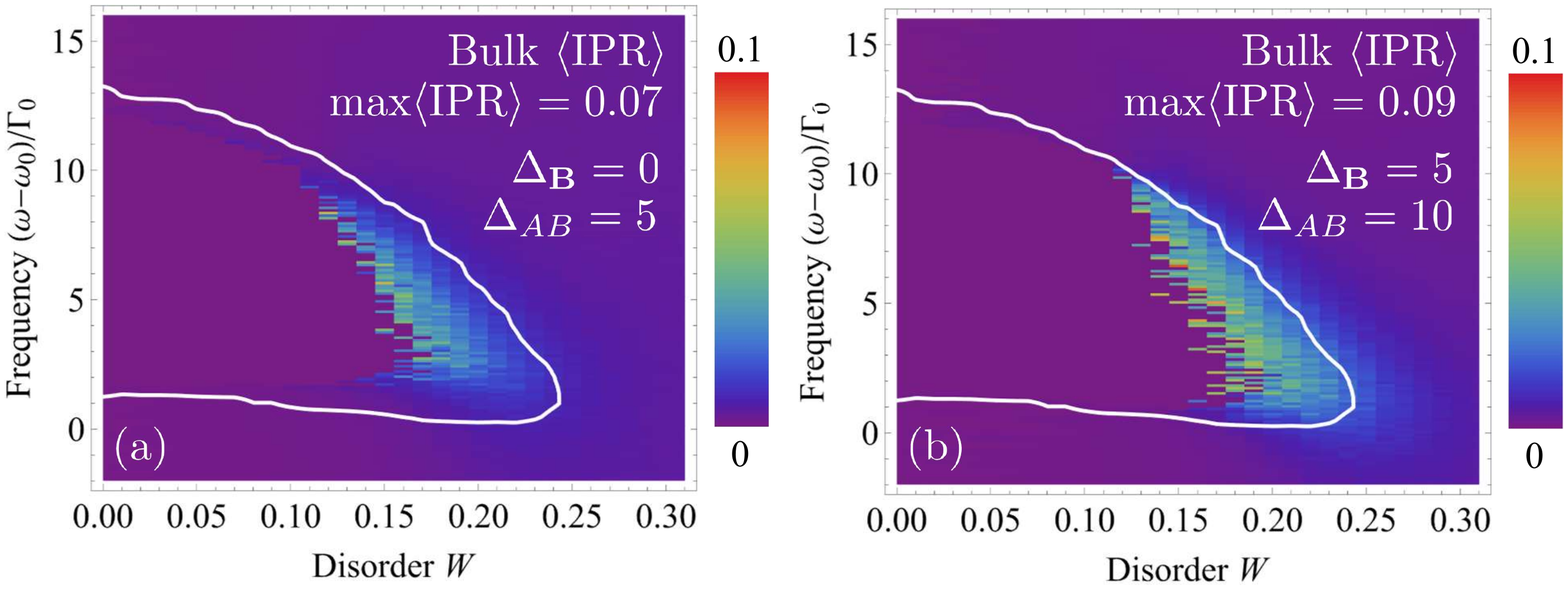}
\vspace{-4.7cm}
\caption{\label{figipr}
Impact of topological properties on spatial localization of modes. Mean bulk IPR in the vicinity of topologically trivial gaps for (a) $\Delta_{\vec{B}} = 0$, $\Delta_{AB} = 5$ and  (b) $\Delta_{\vec{B}} = 5$, $\Delta_{AB} = 10$.
}
\end{figure*}

Figure \ref{fig50} illustrates a topological transition trigged by disorder: the topological band gap where $\langle C_B \rangle = -1$ shrinks upon increasing disorder strength $W$ and closes around $W \simeq 0.25$. By comparing Fig.\ \ref{fig50}(a) with a corresponding plot of bulk density of states (DOS, not shown), we conclude that a contour $\langle C_B \rangle = -0.5$ gives a reasonably accurate description of band edges if we define the latter as frequencies at which DOS drops significantly, albeit not to zero because disorder-induced, spatially localized modes may exist inside the band gap, see Fig.\ \ref{fig50}(c). In addition to disorder-induced modes, discrete topologically protected edge modes with frequencies in the band gap remain resolved in the edge DOS up to $W \simeq 0.15$, see Fig.\ \ref{fig50}(b). The spacing $\delta \omega$ between these modes is inversely proportional to the perimeter of the sample: $\delta \omega \propto 1/\sqrt{N}$, confirming their edge character.

The impact of disorder on the topologically protected edge modes can be understood in the framework of the standard perturbation theory \cite{landau77} generalized to the case of non-Hermitian matrices \cite{sternheim72,moiseyev11}. We represent the Green's matrix $\hat G$ of the disordered lattice as a sum of the unperturbed matrix ${\hat G}^{(0)}$ (eigenvalues $\Lambda^{(0)}_{\alpha}$, right and left eigenvectors $|R^{(0)}_{\alpha}\rangle$ and $|L^{(0)}_{\alpha}\rangle$, respectively) corresponding to the regular lattice and a perturbation $\delta \hat G$ that we assume to be small: ${\hat G} = {\hat G}^{(0)} + {\delta \hat G}$. ${\delta \hat G}$ is obtained by expanding the elements of $\hat G$ in Taylor series in atomic displacements $\Delta \vec{r}_m = \{\Delta x_m, \Delta y_m \}$ up to the second order: $\delta \hat G = (Wa) {\hat V}^{(1)} + (Wa)^2 {\hat V}^{(2)}$.
Explicit expressions for ${\hat V}^{(1)}$ and ${\hat V}^{(2)}$ are given in Appendix \ref{appperturb}.
The average eigenvalues then become
\begin{eqnarray}
\langle \Lambda_{\alpha} \rangle = \Lambda^{(0)}_{\alpha} + (W a) \langle \Lambda^{(1)}_{\alpha} \rangle + (W a)^2 \langle \Lambda^{(2)}_{\alpha} \rangle
\label{perturb}
\end{eqnarray}
with
\begin{eqnarray}
\langle \Lambda^{(1)}_{\alpha} \rangle  &=& \langle L^{(0)}_{\alpha}| \langle \hat V^{(1)} \rangle  |R^{(0)}_{\alpha}\rangle = 0
\label{lambda1av}
\\
\langle \Lambda^{(2)}_{\alpha} \rangle &=& \langle L^{(0)}_{\alpha}| \langle \hat V^{(2)} \rangle |R^{(0)}_{\alpha}\rangle
\nonumber  \\
&+& \sum\limits_{\beta \ne \alpha} \frac{\langle \langle L^{(0)}_{\alpha}| \hat V^{(1)} |R^{(0)}_{\beta}\rangle \langle L^{(0)}_{\beta}| \hat V^{(1)} |R^{(0)}_{\alpha}\rangle \rangle}{\Lambda^{(0)}_{\alpha} - \Lambda^{(0)}_{\beta}}
\label{lambda2av}
\end{eqnarray}
Ensemble averaging denoted by $\langle \ldots \rangle$ can be performed analytically using $\langle \Delta x_m \Delta x_n \rangle = \delta_{mn} (Wa)^2/6$ and  $\langle \Delta x_m \Delta y_n \rangle = 0$, resulting in lengthy but analytical formulas that yield white dashed lines in Fig.\ \ref{fig50}(b).
We refer the reader to Appendix \ref{appperturb} for details of derivations.
The results of the perturbation theory are in remarkable agreement with our numerical calculation as far as the edge modes are well-resolved in DOS. In addition, the frequencies of the lowest- and highest-frequency edge modes provide good approximations for band edges.

The edge modes are not any more well-resolved for $W > 0.15$ but their spectrum always remains well confined within the band gap. The blurring of well-resolved, discrete edge modes roughly coincides with the appearance of spatially localized modes in the bulk of the sample as we illustrate in Fig.\ \ref{fig50}(c) that shows the average inverse participation ratio (IPR) of the modes $| R_{\alpha} \rangle$ in the bulk:
\begin{eqnarray}
\text{IPR}_{\alpha} &=& \frac{\sum_{m=1}^N (\sum_{\sigma = \pm 1} |R_{\alpha m \sigma}|^2)^2}{(\sum_{m=1}^N \sum_{\sigma = \pm 1} |R_{\alpha m \sigma}|^2)^2}
\label{ipr}
\end{eqnarray}
Here $R_{\alpha m \sigma}$ is the amplitude of the $\sigma$-polarized component of the mode $\alpha$ on the atom $m$. Because $\text{IPR} \sim 1/M$ typically corresponds to a mode localized on $M$ atoms, Fig.\ \ref{fig50}(c) evidences that sufficiently strong disorder ($W \gtrsim 0.1$) introduces into the band gap modes that are typically localized on 30--100 atoms.

It is important to remember that the considered physical system is non-Hermitian because atoms can radiate energy into free space surrounding the atomic lattice. As a consequence, the quasimodes have finite lifetimes and decay with a rate $\Gamma_{\alpha} = \Gamma_0 \text{Im} \Lambda_{\alpha}$. Figure \ref{fig50}(d) shows the average decay rate of the quasimodes with frequencies in the topological band gap and in its vicinity. We see that edge modes inside the bandgap have large decay rates (and hence short lifetimes) whereas bulk modes outside the band gap feature slower decays. The decay rate of edge modes becomes independent of the size of the atomic lattice in the limit of large number of atoms $N$, a limit that is already reached for $N$ considered in this work. Difference in lifetimes between bulk and edge modes could be expected since localization at the edge of the system makes leakage of energy to the outside more probable. Short lifetimes of topologically protected edge modes can complicate their observation and practical use.

\begin{figure}
\includegraphics[width=1.1\columnwidth]{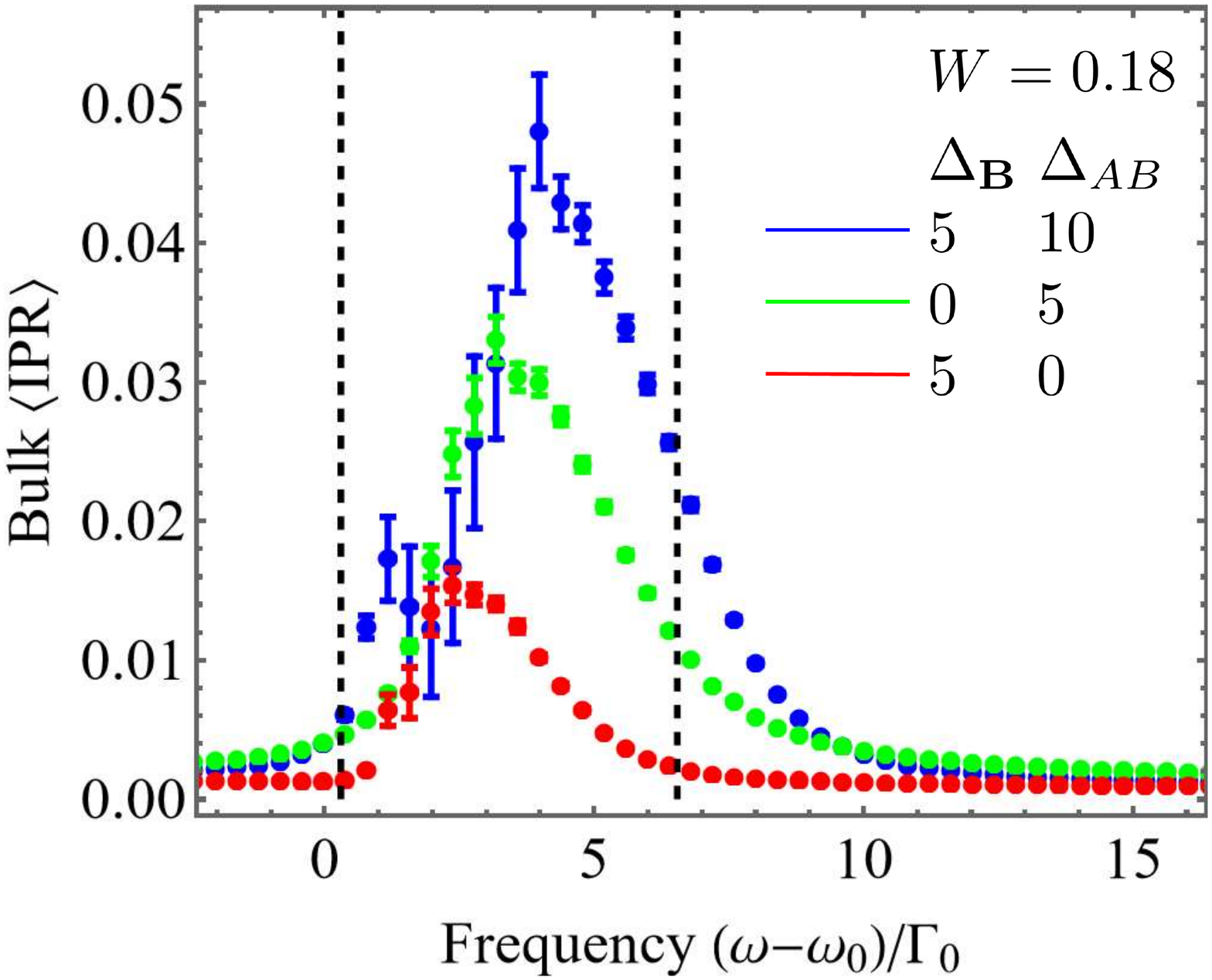}
\caption{\label{figipr18}
Sections of Figs.\ \ref{fig50}(c) (red), \ref{figipr}(a) (green), and \ref{figipr}(b) (blue) at a given disorder $W = 0.18$, averaged over $2000$ realizations of disorder. Error bars show the standard error of the mean. Vertical dashed lines show band gap edges.
}
\end{figure}

Interestingly enough, the degree of localization of modes in the bulk of the disordered system turns out to be affected by the topological nature of the band gap. Indeed, our model of two-level atoms arranged in a honeycomb lattice allows for opening of a band gap of the same width in three different ways: by breaking either the time-reversal ($\Delta_{\mathbf{B}} \ne 0$, $\Delta_{AB} = 0$) or inversion ($\Delta_{\mathbf{B}} = 0$, $\Delta_{AB} \ne 0$) symmetry, or by breaking both of them but to different degrees ($|\Delta_{\mathbf{B}}| \ne |\Delta_{AB}| \ne 0$). The corresponding IPRs are shown in Figs.\ \ref{fig50}(c), \ref{figipr}(a) and \ref{figipr}(b), respectively. In the case of breakdown of both symmetries [Fig.\ \ref{figipr}(b)], the maximum average IPR is roughly a factor of 3 larger than in the case when only the time-reversal symmetry is broken [Fig.\ \ref{fig50}(c)], the situation of broken inversion symmetry being intermediate [Fig.\ \ref{figipr}(a)]. This is also illustrated in Fig.\ \ref{figipr18} where we show $\langle \text{IPR} \rangle$ as a function of frequency at a representative disorder strength $W = 0.18$. Note that IPR in Figs.\ \ref{fig50}(c), \ref{figipr} and \ref{figipr18} is averaged only over bulk modes in order to make the comparison fair and avoid the influence of edge modes that exist in the case of the topological gap only.

When analyzing disorder-induced modes, one should remember that in topologically trivial, fully disordered systems, the breakdown of time-reversal symmetry modifies the symmetry class of the Hamiltonian and generally leads to less pronounced localization and smaller IPR \cite{evers08}. However, this is not what we observe in Figs.\ \ref{fig50}(c), \ref{figipr} and \ref{figipr18} where $\langle \text{IPR} \rangle$ is clearly not controlled by the value of $\Delta_{\mathbf{B}}$ only. Moreover, the largest values of $\langle \text{IPR} \rangle$ and hence the strongest localization are achieved for $\Delta_{\mathbf{B}} = 5$, $\Delta_{AB} = 10$, corresponding to a broken time-reversal symmetry. Our analysis suggests that the topological nature of a band gap reduces IPR and hence increases the localization length $\xi \propto \text{IPR}^{-1/2}$ of disorder-induced modes inside the band gap.

\begin{figure*}
\includegraphics[width=0.8\textwidth]{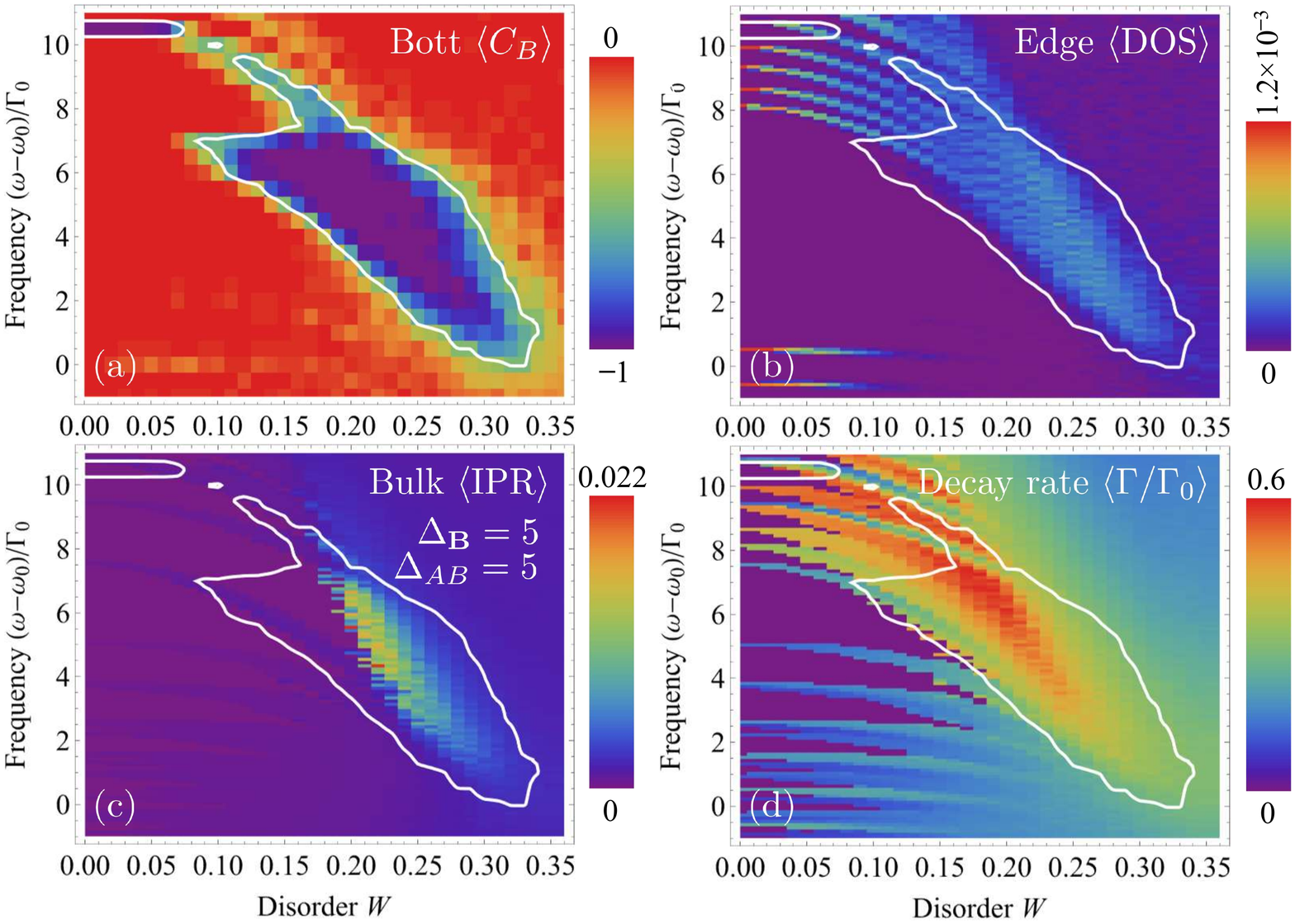}
\caption{\label{fig55}
Opening of a topological pseudogap by disorder. The four plots show the same quantities as Fig.\ \ref{fig50} but for $\Delta_{\vec{B}} = \Delta_{AB} = 5$ and disorder on atoms $B$ only.
}
\end{figure*}

\section{Topological Anderson insulator}
\label{sectai}

In addition to suppression of topological properties by a sufficiently strong disorder, atomic lattices can also exhibit a much less intuitive phenomenon of TAI, in which disorder triggers a transition from a topologically trivial phase to a topologically nontrivial one \cite{li09,groth09,stutzer18,meier18,liu20}. TAI requires both time-reversal ($\Delta_{\mathbf{B}} \ne 0$) and inversion ($\Delta_{AB} \ne 0$) symmetries to be broken.
However, we find that even in such a situation, introducing disorder by displacing all the atoms randomly from their positions in the regular lattice as in Sec.\ \ref{secclosing} does not produce TAI. This topological phase can be reached with the help of disorder that perturbs only one of the two triangular sublattices (the sublattice $B$ here). As far as it remains weak enough, disorder does not modify the physics of the system drastically and,
in the lowest-order approximation, can be incorporated into the Hamiltonian of the periodic lattice by renormalizing its parameters \cite{groth09}. 
The relevant parameters here are $\Delta_{\vec{B}}$ and $\Delta_{AB}$. Disorder makes the effective values of $\Delta_{\vec{B}}$ and $\Delta_{AB}$ deviate from their nominal values and can drive the system into a parameter regime in which a topological band gap would be expected in a regular lattice. The effective $\Delta_{\vec{B}}$ and $\Delta_{AB}$ could be found, for example, from the self-consistent Born approximation \cite{groth09} but the use of the latter is complicated here because of the way in which disorder is introduced (positional disorder). For this reason, we do not implement any analytical calculation here and present only numerical results. Figure \ref{fig55} illustrates the opening of a topological pseudogap for $\Delta_{\mathbf{B}} = \Delta_{AB} = 5$. An ``island'' in the disorder-frequency plane where $\langle C_B \rangle = -1$ is clearly seen in  Fig.\ \ref{fig55}(a). Edge modes are concentrated inside the same island, see Fig.\ \ref{fig55}(b), similarly to what we see in Fig.\ \ref{fig50}(b) for a band gap due to $|\Delta_{\mathbf{B}}| > |\Delta_{AB}|$. We also find localized bulk modes with frequencies in the gap and average IPR up to 0.022, Fig.\ \ref{fig55}(c). Note that here both the opening of the pseudogap and the appearance of localized modes inside it are due to disorder in contrast to Fig.\ \ref{fig50} where the gap exists in the absence of disorder as well. Localized modes arise at a finite disorder strength [$W \simeq 0.2$ in Fig.\ \ref{fig55}(c)] and disappear when disorder is too strong (for $W \gtrsim 0.3$). This is different from the standard Anderson localization scenario in which all modes in a ceratin frequency band would be localized for all $W$ exceeding some critical $W_c$, with a possibility of $W_c = 0$ in 1D and 2D \cite{evers08}.

Similarly to the case of the topological band gap in Fig.\ \ref{fig50}, decay rates of quasimodes with frequencies inside the topological band gap in Fig.\ \ref{fig55} are relatively high, see Fig.\ \ref{fig55}(d). Again, this is due to their localization at the boundaries of the sample.

\section{Discussion}
\label{secdisc}

Let is now discuss some unexpected features of our results. An attentive reader may note that both Figs.\ \ref{fig50}(b) and \ref{fig55}(b) exhibit edge modes extending outside of the topological band gap delimited by the white contours $\langle C_B \rangle = -0.5$. We explored the spatial structure of these modes and found that they are due to two distinct phenomena. First, it turns out that the spectral range in which the external magnetic field introduces edge modes is not restricted to the band gap. Such modes also arise outside the bandgap, although in the immediate vicinity of band edges. They are regularly spaced in frequency and clearly visible in Fig.\ \ref{fig50}(b). However, in contrast to edge modes inside the topological band gap, edge modes outside the band gap are accompanied by bulk modes with similar frequencies. Thus, even if they are clearly visible, edge modes with frequencies outside the band gap do not benefit from the topological protection since any perturbation can easily couple them to bulk modes of arbitrarily close frequencies. For this reason, they are visible only for $W \lesssim 0.1$ in Fig.\ \ref{fig50}(b) whereas the modes inside the band gap persist until $W \simeq 0.25$ even though their eigenfrequencies are washed out by disorder at large $W$. 

A second phenomenon that leads to appearance of edge modes with frequencies outside of topological band gaps is due to the corners of our hexagon-shaped samples. The arm-chair symmetry of edges is broken at the corners, which turns out to be particularly important under conditions of TAI (Fig.\ \ref{fig55}): the six edge modes visible at $(\omega-\omega_0)/\Gamma_0 \gtrsim 8$, $W \lesssim 0.1$ have strong maxima at the corners of the sample. These ``corner states'' exist already in the regular lattice ($W = 0$) and survive until the opening of the topological band gap at $W \simeq 0.1$. At stronger disorder, corners of the lattice cease to play a special role and edge modes do not tend to get localized around them anymore.  Another artifact due to a mode that is strongly localized in a corner of the square sample (used for the calculation of $C_B$), is a narrow frequency band with $\langle C_B \rangle = -1$ around $(\omega-\omega_0)/\Gamma_0 \simeq 10.5$ in Fig.\ \ref{fig55}(a). It remains to be seen whether the existence of corner modes in our system is an indication of higher-order topological insulator physics \cite{benalcazar17,schindler18}. A study of this question is beyond the scope of the present work.

\section{Conclusions}
\label{secconcl}

In conclusion, we reveal the mutual impact of disorder and topology in atomic lattices. On the one hand, disorder can trigger transitions between topologically distinct phases and, in particular, drive the lattice into the optical TAI phase. On the other hand, the type of topological phase influences the degree of disorder-induced spatial localization of modes. On average, IPR of most localized modes with frequencies inside a topological band gap can be a factor of 3 smaller than IPR of modes inside a trivial band gap. A particularly interesting phenomenon in which both disorder and topology are involved is the appearance of disorder-induced spatially localized bulk modes assisted by the opening of a topological pseudogap induced by disorder as well.

\acknowledgements
The authors thank Fabrice Mortessagne and Bart van Tiggelen for useful discussions. This work was funded by the Agence Nationale de la Recherche (Grant No. ANR-20-CE30-0003 LOLITOP).

\appendix

\section{Calculation of Bott index}
\label{appbott}

Let us introduce a basis of localized states $|n, \sigma \rangle$ ($n = 1, \ldots, N$, $\sigma = \pm 1$). A state $|n, \sigma \rangle$ is localized on the atom $n$ and have the polarization $\sigma$. We expand right and left eigenvectors of the matrix $\hat{G}$ over this basis: 
\begin{eqnarray}
| R_{\alpha} \rangle  &=& \sum\limits_{n = 1}^{N} \sum\limits_{\sigma = \pm 1} R_{\alpha n \sigma} | n, \sigma \rangle
\label{expansionR}
\\
\langle L_{\alpha} |  &=& \sum\limits_{n = 1}^{N} \sum\limits_{\sigma = \pm 1} L_{\alpha n \sigma}^* \langle n, \sigma |
\label{expansionL}
\end{eqnarray}
where $R_{\alpha n \sigma}$ and $ L_{\alpha n \sigma}^*$ are weights of $\sigma$-components of the corresponding eigenvectors on the atom $n$.

The matrices ${\hat V}_{X,Y}$ defined by Eq.\ (\ref{vxy}) can be written as
\begin{eqnarray}
{\hat V}_{X,Y}  &=& {\hat P} {\hat U}_{X,Y} {\hat P}
\nonumber \\
&=& \sum\limits_{\omega_{\alpha},\omega_{\beta} \leq \omega} | R_{\alpha} \rangle \langle L_{\alpha}| {\hat U}_{X,Y} | R_{\beta} \rangle \langle L_{\beta}|
\nonumber \\
&=& \sum\limits_{\omega_{\alpha},\omega_{\beta} \leq \omega} | R_{\alpha} \rangle ( {\hat V}_{X,Y} )_{\alpha \beta} \langle L_{\beta}|
\label{utildeRL}
\end{eqnarray}
where
\begin{eqnarray}
( {\hat V}_{X,Y} )_{\alpha \beta} &=& \langle L_{\alpha}| {\hat U}_{X,Y} | R_{\beta} \rangle
\nonumber \\
&=& \sum\limits_{m,n = 1}^{N}  \sum\limits_{\sigma,\rho = \pm 1}
L_{\alpha m \sigma}^* \langle m, \sigma | {\hat U}_{X,Y} | n, \rho \rangle
R_{\beta n \rho}
\nonumber \\
&=& \sum\limits_{n = 1}^{N} \sum\limits_{\sigma = \pm 1}
L_{\alpha n \sigma}^* (\hat{U}_{X,Y})_{nn} R_{\beta n \sigma}
\label{uabRL}
\end{eqnarray}
By noting that $({\hat U}_{X})_{nn} =  \exp(i 2\pi x_n/L_x)$ [and similarly for $({\hat U}_{Y})_{nn}$], we can rewrite Eq.\ (\ref{uabRL}) for the elements $\alpha, \beta$ of the matrices ${\hat V}_{X,Y}$ as
\begin{eqnarray}
( {\hat V}_{X} )_{\alpha \beta} &=&
\sum\limits_{n = 1}^{N} \sum\limits_{\sigma = \pm1} 
L_{\alpha n \sigma}^* R_{\beta n \sigma}
e^{i 2\pi x_n/L_x}
\label{umnfinalxRL}
\\
( {\hat V}_{Y} )_{\alpha \beta} &=&
\sum\limits_{n = 1}^{N} \sum\limits_{\sigma = \pm1} 
L_{\alpha n \sigma}^* R_{\beta n \sigma}
e^{i 2\pi y_n/L_y}
\label{umnfinalyRL}
\end{eqnarray}
The Bott index can now be evaluated from Eq.\ (\ref{bott}):
\begin{eqnarray}
C_B &=&  \frac{1}{2\pi} \mathrm{Im} \mathrm{Tr} \ln [{\hat V}_X {\hat V}_Y {\hat V}_X^{\dagger} {\hat V}_Y^{\dagger}]
\nonumber \\
&=& \frac{1}{2\pi} \mathrm{Im} \mathrm{Tr} \ln {\hat W}
= \frac{1}{2\pi} \mathrm{Im} \ln \det {\hat W}
\nonumber \\
&=& \frac{1}{2\pi} \mathrm{Im} \ln \prod\limits_n W_n
= \frac{1}{2\pi} \sum\limits_n \mathrm{Im} \left( \ln W_n \right)
\label{bott2}
\end{eqnarray}
where 
${\hat W} = {\hat V}_X {\hat V}_Y {\hat V}_X^{\dagger} {\hat V}_Y^{\dagger}$ and  $W_n$ are eigenvalues of ${\hat W}$.
Because $R_{\alpha n \sigma}$ and $L_{\alpha n \sigma}^*$ are direct results of any numerical algorithm for funding eigenvectors of the matrix $\hat{G}$ (such as, e.g., {\tt zgeev} from LAPACK library \cite{lapack}),  Eqs.\ (\ref{umnfinalxRL}), (\ref{umnfinalyRL}) and (\ref{bott2}) allow us to evaluate $C_B$ numerically. After averaging over many different realizations of disorder, they yield Figs.\ \ref{fig50}(a) and  \ref{fig55}(a) of the main text.

Note that the size of matrices ${\hat V}_{X,Y}$ and ${\hat W}$ as well as the number of eigenvalues $W_n$ are equal to the number of eigenfrequencies $\omega_{\alpha}$ below $\omega$. It is the only way in which ${\hat V}_{X,Y}$, ${\hat W}$ and $C_B$ depend on $\omega$.

\section{Perturbation theory for the spectrum of edge states}
\label{appperturb}

We represent the Green's matrix $\hat G$ of the disordered lattice as a sum of the unperturbed matrix ${\hat G}^{(0)}$ corresponding to the regular lattice and a perturbation $\delta \hat G$ that we assume to be small:
\begin{eqnarray}
{\hat G} &=& {\hat G}^{(0)} + {\delta \hat G}
\label{perturbapp}
\end{eqnarray}
The perturbation ${\delta \hat G}$ is obtained by expanding the elements of $\hat G$ in Taylor series in atomic displacements $\Delta \vec{r}_m = \{\Delta x_m, \Delta y_m \}$:
\begin{eqnarray}
{\delta G}_{mn} &=& \frac{\partial G^{(0)}_{mn}}{\partial(\xi_m - \xi_n)} (\Delta \xi_m - \Delta \xi_n)
\nonumber \\
&+& \frac{\partial G^{(0)}_{mn}}{\partial(\eta_m - \eta_n)} (\Delta \eta_m - \Delta \eta_n)
\nonumber \\
&+& \frac12 \frac{\partial^2 G^{(0)}_{mn}}{\partial(\xi_m - \xi_n)^2} (\Delta \xi_m - \Delta \xi_n)^2
\nonumber \\
&+& \frac12 \frac{\partial^2 G^{(0)}_{mn}}{\partial(\eta_m - \eta_n)^2} (\Delta \eta_m - \Delta \eta_n)^2
\nonumber \\
&+& \frac{\partial^2 G^{(0)}_{mn}}{\partial(\xi_m - \xi_n)\partial(\eta_m - \eta_n)}
\nonumber \\
&\times& (\Delta \xi_m - \Delta \xi_n)(\Delta \eta_m - \Delta \eta_n)
\label{taylor}
\end{eqnarray}
where we introduced vectors
\begin{eqnarray}
\bm{\xi} &=& \{x_1, x_1, x_2, x_2, \ldots, x_N, x_N \}
\label{vecxi}\\
\bm{\eta} &=& \{y_1, y_1, y_2, y_2, \ldots, y_N, y_N \}
\label{veceta}
\end{eqnarray}
of length $2N$ in order to have all vectors and matrices of the same dimension ($2N$ and $2N \times 2N$, respectively). We keep terms up to the second order in $\Delta \vec{r}_m$ because we aim at a result that is second order in disorder strength $W$.

Introducing dimensionless displacements $\Delta  \tilde{\vec{r}}_m = \Delta \vec{r}_m/(Wa)$ that vary between 0 and 1 in absolute values, we can write $\delta \hat G$ as
\begin{eqnarray}
\delta \hat G = (Wa) {\hat V}^{(1)} + (Wa)^2 {\hat V}^{(2)}
\label{taylor2}
\end{eqnarray}
where the elements of matrices ${\hat V}^{(1)}$ and ${\hat V}^{(2)}$ are
\begin{eqnarray}
V^{(1)}_{mn} &=& 
D^{(\xi)}_{mn} (\Delta {\tilde \xi}_m - \Delta {\tilde \xi}_n)
+ D^{(\eta)}_{mn} (\Delta {\tilde \eta}_m - \Delta {\tilde \eta}_n)
\\
V^{(2)}_{mn} &=&
\frac12 D^{(\xi \xi)}_{mn} (\Delta {\tilde \xi}_m - \Delta {\tilde \xi}_n)^2
+   \frac12 D^{(\eta \eta)}_{mn} (\Delta {\tilde \eta}_m - \Delta {\tilde \eta}_n)^2
\nonumber \\
&+& D^{(\xi \eta)}_{mn}
(\Delta {\tilde \xi}_m - \Delta {\tilde \xi}_n)(\Delta {\tilde \eta}_m - \Delta {\tilde \eta}_n)
\label{taylor3}
\end{eqnarray}
where, in order to make notation more compact, we introduced $2N \times 2N$ derivative matrices
\begin{eqnarray}
D^{(\xi)}_{mn} &=& \frac{\partial G^{(0)}_{mn}}{\partial(\xi_m - \xi_n)}
\label{bxi} \\
D^{(\eta)}_{mn} &=& \frac{\partial G^{(0)}_{mn}}{\partial(\eta_m - \eta_n)}
\label{beta} \\
D^{(\xi \xi)}_{mn} &=& \frac{\partial^2 G^{(0)}_{mn}}{\partial(\xi_m - \xi_n)^2}
\label{bxixi} \\
D^{(\eta \eta)}_{mn} &=& \frac{\partial^2 G^{(0)}_{mn}}{\partial(\eta_m - \eta_n)^2}
\label{betaeta} \\
D^{(\xi \eta)}_{mn} &=& \frac{\partial^2 G^{(0)}_{mn}}{\partial(\xi_m - \xi_n)\partial(\eta_m - \eta_n)}
\label{bxieta}
\end{eqnarray}

We now denote by $\Lambda^{(0)}_{\alpha}$ ($\tilde\Lambda^{(0)}_{\alpha}$) and $|R^{(0)}_{\alpha}\rangle$ ($|L^{(0)}_{\alpha}\rangle$) the right (left) eigenvalues and eigenvectors of the matrix ${\hat G}^{(0)}$. In the nondegenerate case when all eigenvalues are different, the eigenvectors can be normalized to form a biorthogonal basis: $\langle L^{(0)}_{\alpha} | R^{(0)}_{\beta}\rangle = \delta_{\alpha \beta}$.
The perturbation theory allows us to compute corrections to the unperturbed eigenvalues $\Lambda^{(0)}_{\alpha}$ in successive orders of the small parameter $Wa \ll 1$ \cite{sternheim72}. Keeping only terms up to order $(Wa)^2$ we get
\begin{eqnarray}
\Lambda_{\alpha} &=& \Lambda^{(0)}_{\alpha} + (W a) \Lambda^{(1)}_{\alpha} + (W a)^2 \Lambda^{(2)}_{\alpha}
\label{lambda}
\\
\Lambda^{(1)}_{\alpha} &=& \langle L^{(0)}_{\alpha}| \hat V^{(1)} |R^{(0)}_{\alpha}\rangle
\label{lambda1}
\\
\Lambda^{(2)}_{\alpha} &=& \langle L^{(0)}_{\alpha}| \hat V^{(2)} |R^{(0)}_{\alpha}\rangle
\nonumber  \\
&+& \sum\limits_{\beta \ne \alpha} \frac{\langle L^{(0)}_{\alpha}| \hat V^{(1)} |R^{(0)}_{\beta}\rangle \langle L^{(0)}_{\beta}| \hat V^{(1)} |R^{(0)}_{\alpha}\rangle}{\Lambda^{(0)}_{\alpha} - \Lambda^{(0)}_{\beta}}
\label{lambda2}
\end{eqnarray}

Averaging of Eqs.\ (\ref{lambda1}) and (\ref{lambda2}) over disorder yields
\begin{eqnarray}
\langle \Lambda^{(1)}_{\alpha} \rangle &=& \langle L^{(0)}_{\alpha}| \langle \hat V^{(1)} \rangle |R^{(0)}_{\alpha}\rangle = 0
\label{lambda1avapp}
\\
\langle \Lambda^{(2)}_{\alpha} \rangle &=& \langle L^{(0)}_{\alpha}| \langle \hat V^{(2)} \rangle |R^{(0)}_{\alpha}\rangle
\nonumber  \\
&+& \sum\limits_{\beta \ne \alpha} \frac{\langle \langle L^{(0)}_{\alpha}| \hat V^{(1)} |R^{(0)}_{\beta}\rangle \langle L^{(0)}_{\beta}| \hat V^{(1)} |R^{(0)}_{\alpha}\rangle \rangle}{\Lambda^{(0)}_{\alpha} - \Lambda^{(0)}_{\beta}}\;\;\;\;\;\;
\label{lambda2avapp}
\end{eqnarray}
which are Eqs.\ (\ref{lambda1av}) and (\ref{lambda2av}) of the main text. The first equation follows from the isotropy of disorder: $\langle \Delta x_n \rangle = \langle \Delta y_n \rangle = 0$. To evaluate the first term on the right-hand side (r.h.s.) of Eq.\ (\ref{lambda2avapp}) we note that
\begin{eqnarray}
\langle V^{(2)}_{mn} \rangle &=&
\frac12 D^{(\xi \xi)}_{mn} \langle (\Delta {\tilde \xi}_m - \Delta {\tilde \xi}_n)^2 \rangle
\nonumber \\
&+& \frac12 D^{(\eta \eta)}_{mn} \langle (\Delta {\tilde \eta}_m - \Delta {\tilde \eta}_n)^2 \rangle
\nonumber \\
&+& D^{(\xi \eta)}_{mn} \langle (\Delta {\tilde \xi}_m - \Delta {\tilde \xi}_n)(\Delta {\tilde \eta}_m - \Delta {\tilde \eta}_n) \rangle
\nonumber \\
&=& \frac16 \left[ D^{(\xi \xi)}_{mn} + D^{(\eta \eta)}_{mn} \right]
\label{lambda2av1}
\end{eqnarray}
To obtain the last line of this equation, we used $\langle \Delta {\tilde \xi}_m^2 \rangle = \langle \Delta {\tilde \eta}_m^2 \rangle = 1/6$ and $\langle \Delta {\tilde \xi}_m \Delta {\tilde \eta}_n \rangle = 0$. Note that $\Delta {\tilde \xi}_m - \Delta {\tilde \xi}_n = \Delta {\tilde \eta}_m - \Delta {\tilde \eta}_n = 0$ when the indices $m$ and $n$ correspond to the same atom ($n = m$ and $m+1$ for odd $m$ or $n = m$ and $m-1$ for even $m$), suggesting that this case should be treated separately. However, the derivatives in Eq.\ (\ref{lambda2av1}) also vanish in such a situation, permitting us to extend Eq.\ (\ref{lambda2av1}) to all $m, n = 1, \ldots, 2N$.

The second term on the r.h.s. of  Eq.\ (\ref{lambda2avapp}) requires evaluation of the following average:
\begin{eqnarray}
&&\langle \langle L^{(0)}_{\alpha}| \hat V^{(1)} |R^{(0)}_{\beta}\rangle \langle L^{(0)}_{\beta}| \hat V^{(1)} |R^{(0)}_{\alpha}\rangle \rangle
\nonumber \\
&&= \sum\limits_{i,j,k,l = 1}^{2N} \langle L^{(0)*}_{\alpha i } V^{(1)}_{ij} R^{(0)}_{\beta j} L^{(0)*}_{\beta k} V^{(1)}_{kl} R^{(0)}_{\beta l} \rangle
\nonumber \\
&&= \sum\limits_{i,j,k,l = 1}^{2N} L^{(0)*}_{\alpha i } R^{(0)}_{\beta j} L^{(0)*}_{\beta k} R^{(0)}_{\beta l} \langle V^{(1)}_{ij} V^{(1)}_{kl} \rangle
\label{lambda2av2}
\end{eqnarray}
where we use a notation
$R_{\alpha i}^{(0)} = R_{\alpha n \sigma}^{(0)}$ with $i = 2n + (\sigma-1)/2$ (and similarly for $L_{\alpha i}^{(0)*}$). The disorder average in this equation equals to
\begin{eqnarray}
\langle V^{(1)}_{ij} V^{(1)}_{kl} \rangle &=&
D^{(\xi)}_{ij} D^{(\xi)}_{kl} \langle (\Delta {\tilde \xi}_i - \Delta {\tilde \xi}_j)(\Delta {\tilde \xi}_k - \Delta {\tilde \xi}_l) \rangle
\nonumber \\
&+& D^{(\eta)}_{ij} D^{(\eta)}_{kl} \langle (\Delta {\tilde \eta}_i - \Delta {\tilde \eta}_j) (\Delta {\tilde \eta}_k - \Delta {\tilde \eta}_l) \rangle
\nonumber \\
&=& D^{(\xi)}_{ij} D^{(\xi)}_{kl} \left(
\langle \Delta {\tilde \xi}_i \Delta {\tilde \xi}_k \rangle
+ \langle \Delta {\tilde \xi}_j \Delta {\tilde \xi}_l \rangle
\right.
\nonumber \\
&-& \left. \langle \Delta {\tilde \xi}_i \Delta {\tilde \xi}_l \rangle
- \langle \Delta {\tilde \xi}_j \Delta {\tilde \xi}_k \rangle
\right)
\nonumber \\
&+& D^{(\eta)}_{ij} D^{(\eta)}_{kl} \left(
\langle \Delta {\tilde \eta}_i \Delta {\tilde \eta}_k \rangle
+ \langle \Delta {\tilde \eta}_j \Delta {\tilde \eta}_l \rangle
\right.
\nonumber \\
&-& \left. \langle\Delta {\tilde \eta}_i \Delta {\tilde \eta}_l \rangle
- \langle \Delta {\tilde \eta}_j \Delta {\tilde \eta}_k \rangle
\right)
\nonumber \\
&=& \frac16 \left( D^{(\xi)}_{ij} D^{(\xi)}_{kl} + D^{(\eta)}_{ij} D^{(\eta)}_{kl} \right)
\nonumber \\
&\times& \left(
\Delta_{ik}
+ \Delta_{jl}
- \Delta_{il}
- \Delta_{jk}
\right)
\label{v1v1}
\end{eqnarray}
where, again, we used the statistical independence of $\Delta {\tilde \xi}_m$ and $\Delta {\tilde \eta}_n$: $\langle \Delta {\tilde \xi}_m \Delta {\tilde \eta}_n \rangle = 0$,
and the fact that $\langle\Delta {\tilde \xi}_i \Delta {\tilde \xi}_k \rangle = \langle \Delta {\tilde \eta}_i \Delta {\tilde \eta}_k \rangle = \Delta_{ik}/6$. Here we introduce a modified ``Kronecker symbol'' $\Delta_{ik}$ that equals to either 1 when the indices $i$ and $k$ correspond to the same atom or 0 in the opposite case:
\begin{eqnarray}
\Delta_{ij} =
\begin{cases}
1, &j = i \text{ or } j = \tilde i\\
0, &\text{otherwise}
\end{cases}
\label{kronecker}
\end{eqnarray}
where $\tilde i = i + (-1)^{i+1}$

We are now ready to rewrite Eq.\ (\ref{lambda2avapp}) in a form suitable for numerical evaluation. The first term on the r.h.s. of this equation is
\begin{eqnarray}
\langle L^{(0)}_{\alpha}| \langle \hat V^{(2)} \rangle |R^{(0)}_{\alpha}\rangle &=&
\sum\limits_{mn} L^{(0)*}_{\alpha m} \langle V^{(2)}_{mn} \rangle R^{(0)}_{\alpha n}
\nonumber \\
&=& \sum\limits_{mn} \left( L^{(0)*} \right)_{\alpha m} \langle V^{(2)}_{mn} \rangle \left( R^{(0)T} \right)_{n \alpha}
\nonumber \\
&=& \left( {\hat L}^{(0)*} \langle \hat V^{(2)} \rangle {\hat R}^{(0)T} \right)_{\alpha\alpha}
\nonumber \\
&=& \frac16 \left[ {\hat L}^{(0)*} 
\left(
{\hat D}^{(\xi \xi)} + {\hat D}^{(\eta \eta)}
\right)
{\hat R}^{(0)T} \right]_{\alpha\alpha}
\nonumber \\
\label{lambda2av1a}
\end{eqnarray}
where we introduced matrices $\hat R^{(0)}$ and $\hat L^{(0)}$ containing eigenvectors $|R^{(0)}_{\alpha} \rangle$ and  $|L^{(0)}_{\alpha} \rangle$ as rows.

The average in the second term on the r.h.s. of Eq.\ (\ref{lambda2avapp}) is
\begin{eqnarray}
&&\langle \langle L^{(0)}_{\alpha}| \hat V^{(1)} |R^{(0)}_{\beta}\rangle \langle L^{(0)}_{\beta}| \hat V^{(1)} |R^{(0)}_{\alpha}\rangle \rangle
\nonumber \\
&&= \sum\limits_{i,j,k,l = 1}^{2N} L^{(0)*}_{\alpha i } R^{(0)}_{\beta j} L^{(0)*}_{\beta k} R^{(0)}_{\alpha l} \langle V^{(1)}_{ij} V^{(1)}_{kl} \rangle
\nonumber \\
&&= \frac16 \sum\limits_{i,j,k,l = 1}^{2N} L^{(0)*}_{\alpha i } R^{(0)}_{\beta j} L^{(0)*}_{\beta k} R^{(0)}_{\alpha l}
\left( D^{(\xi)}_{ij} D^{(\xi)}_{kl} + D^{(\eta)}_{ij} D^{(\eta)}_{kl} \right)
\nonumber \\
&&\times \left(
\Delta_{ik}
+ \Delta_{jl}
- \Delta_{il}
- \Delta_{jk}
\right)
\nonumber \\
&&=\frac16 \sum\limits_{i,j,l = 1}^{2N} L^{(0)*}_{\alpha i } R^{(0)}_{\beta j} L^{(0)*}_{\beta i} R^{(0)}_{\alpha l}
\left( D^{(\xi)}_{ij} D^{(\xi)}_{il} + D^{(\eta)}_{ij} D^{(\eta)}_{il} \right)
\nonumber \\
&&+
\frac16 \sum\limits_{i,j,l = 1}^{2N} L^{(0)*}_{\alpha i } R^{(0)}_{\beta j} L^{(0)*}_{\beta \tilde i} R^{(0)}_{\alpha l}
\left( D^{(\xi)}_{ij} D^{(\xi)}_{\tilde il} + D^{(\eta)}_{ij} D^{(\eta)}_{\tilde il} \right)
\nonumber \\
&&+ \frac16 \sum\limits_{i,j,k = 1}^{2N} L^{(0)*}_{\alpha i } R^{(0)}_{\beta j} L^{(0)*}_{\beta k} R^{(0)}_{\alpha j}
\left( D^{(\xi)}_{ij} D^{(\xi)}_{kj} + D^{(\eta)}_{ij} D^{(\eta)}_{kj} \right)
\nonumber
\end{eqnarray}
\begin{eqnarray}
&&+
\frac16 \sum\limits_{i,j,k = 1}^{2N} L^{(0)*}_{\alpha i } R^{(0)}_{\beta j} L^{(0)*}_{\beta k} R^{(0)}_{\alpha \tilde j}
\left( D^{(\xi)}_{ij} D^{(\xi)}_{k \tilde j} + D^{(\eta)}_{ij} D^{(\eta)}_{k \tilde j} \right)
\nonumber \\
&&- \frac16 \sum\limits_{i,j,k = 1}^{2N} L^{(0)*}_{\alpha i } R^{(0)}_{\beta j} L^{(0)*}_{\beta k} R^{(0)}_{\alpha i}
\left( D^{(\xi)}_{ij} D^{(\xi)}_{ki} + D^{(\eta)}_{ij} D^{(\eta)}_{ki} \right)
\nonumber \\
&&- \frac16 \sum\limits_{i,j,k = 1}^{2N} L^{(0)*}_{\alpha i } R^{(0)}_{\beta j} L^{(0)*}_{\beta k} R^{(0)}_{\alpha \tilde i}
\left( D^{(\xi)}_{ij} D^{(\xi)}_{k \tilde i} + D^{(\eta)}_{ij} D^{(\eta)}_{k \tilde i} \right)
\nonumber \\
&&- \frac16 \sum\limits_{i,j,l = 1}^{2N} L^{(0)*}_{\alpha i } R^{(0)}_{\beta j} L^{(0)*}_{\beta j} R^{(0)}_{\alpha l}
\left( D^{(\xi)}_{ij} D^{(\xi)}_{jl} + D^{(\eta)}_{ij} D^{(\eta)}_{jl} \right)
\nonumber \\
&&- \frac16 \sum\limits_{i,j,l = 1}^{2N} L^{(0)*}_{\alpha i } R^{(0)}_{\beta j} L^{(0)*}_{\beta \tilde j} R^{(0)}_{\alpha l}
\left( D^{(\xi)}_{ij} D^{(\xi)}_{\tilde j l} + D^{(\eta)}_{ij} D^{(\eta)}_{\tilde j l} \right)
\nonumber \\
\label{lambda2av2a}
\end{eqnarray}

Let us first consider odd terms of in Eq.\ (\ref{lambda2av2a}), i.e. the terms that contain neither $\tilde i$ nor $\tilde j$. The first term can be rewritten as
\begin{eqnarray}
&&\frac16 \sum\limits_{i,j,l = 1}^{2N} L^{(0)*}_{\alpha i } R^{(0)}_{\beta j} L^{(0)*}_{\beta i} R^{(0)}_{\alpha l}
\left( D^{(\xi)}_{ij} D^{(\xi)}_{il} + D^{(\eta)}_{ij} D^{(\eta)}_{il} \right)
\nonumber \\
&& = \frac16 \sum\limits_{i = 1}^{2N} L^{(0)*}_{\alpha i }  L^{(0)*}_{\beta i}
\left[ \left( \sum\limits_{j=1}^{2N} D^{(\xi)}_{ij} R^{(0)}_{\beta j} \right)
\left( \sum\limits_{l=1}^{2N} D^{(\xi)}_{il} R^{(0)}_{\alpha l} \right)
\right.
\nonumber \\
&&+\left. \left( \sum\limits_{j=1}^{2N} D^{(\eta)}_{ij} R^{(0)}_{\beta j} \right)
\left( \sum\limits_{l=1}^{2N} D^{(\eta)}_{il} R^{(0)}_{\alpha l} \right) \right]
\nonumber \\
&&= \frac16 \sum\limits_{i = 1}^{2N} L^{(0)*}_{\alpha i }  L^{(0)*}_{\beta i}
\left[ \left( \hat D^{(\xi)} \hat R^{(0)T} \right)_{i \beta}
\left( \hat D^{(\xi)} \hat R^{(0)T} \right)_{i \alpha}
\right.
\nonumber \\
&&+ \left. \left( \hat D^{(\eta)} \hat R^{(0)T} \right)_{i \beta}
\left( \hat D^{(\eta)} \hat R^{(0)T} \right)_{i \alpha} \right]
\label{term1}
\end{eqnarray}
The third term is
\begin{eqnarray}
&&\frac16 \sum\limits_{i,j,k = 1}^{2N} L^{(0)*}_{\alpha i } R^{(0)}_{\beta j} L^{(0)*}_{\beta k} R^{(0)}_{\alpha j}
\left( D^{(\xi)}_{ij} D^{(\xi)}_{kj} + D^{(\eta)}_{ij} D^{(\eta)}_{kj} \right)
\nonumber \\
&& =\frac16 \sum\limits_{j = 1}^{2N} R^{(0)}_{\beta j} R^{(0)}_{\alpha j}
\left[ \left( \sum\limits_i L^{(0)*}_{\alpha i } D^{(\xi)}_{ij} \right) \left( \sum\limits_k  L^{(0)*}_{\beta k}  D^{(\xi)}_{kj} \right)
\right.
\nonumber \\
&&+ \left.
\left( \sum\limits_i L^{(0)*}_{\alpha i } D^{(\eta)}_{ij} \right) \left( \sum\limits_k  L^{(0)*}_{\beta k}  D^{(\eta)}_{kj} \right) \right]
\nonumber \\
&& =\frac16 \sum\limits_{j = 1}^{2N} R^{(0)}_{\beta j} R^{(0)}_{\alpha j}
\left[ \left( \hat L^{(0)*} \hat D^{(\xi)} \right)_{\alpha j} \left( \hat L^{(0)*} \hat D^{(\xi)} \right)_{\beta j}
\right.
\nonumber \\
&&+ \left. \left( \hat L^{(0)*} \hat D^{(\eta)} \right)_{\alpha j} \left( \hat L^{(0)*} \hat D^{(\eta)} \right)_{\beta j} \right]
\label{term3}
\end{eqnarray}
The firth term of Eq.\ (\ref{lambda2av2a}) is
\begin{eqnarray}
&&\frac16 \sum\limits_{i,j,k = 1}^{2N} L^{(0)*}_{\alpha i } R^{(0)}_{\beta j} L^{(0)*}_{\beta k} R^{(0)}_{\alpha i}
\left( D^{(\xi)}_{ij} D^{(\xi)}_{ki} + D^{(\eta)}_{ij} D^{(\eta)}_{ki} \right) =
\nonumber \\
&&= \frac16 \sum\limits_{i = 1}^{2N} L^{(0)*}_{\alpha i }  R^{(0)}_{\alpha i}
\left[ \left( \sum\limits_j D^{(\xi)}_{ij} R^{(0)}_{\beta j} \right) \left( \sum\limits_{k} L^{(0)*}_{\beta k} D^{(\xi)}_{ki} \right)
\right.
\nonumber \\
&&+\left. \left( \sum\limits_j D^{(\eta)}_{ij} R^{(0)}_{\beta j} \right) \left( \sum\limits_{k} L^{(0)*}_{\beta k} D^{(\eta)}_{ki} \right) \right]
\nonumber \\
&&= \frac16 \sum\limits_{i = 1}^{2N} L^{(0)*}_{\alpha i }  R^{(0)}_{\alpha i}
\left[ \left( \hat D^{(\xi)} R^{(0)T} \right)_{i \beta} \left( \hat L^{(0)*} \hat D^{(\xi)} \right)_{\beta i}
\right.
\nonumber \\
&&+ \left. \left( \hat D^{(\eta)} R^{(0)T} \right)_{i \beta} \left( \hat L^{(0)*} \hat D^{(\eta)} \right)_{\beta i} \right]
\label{term5}
\end{eqnarray}
And the seventh term is
\begin{eqnarray}
&&\frac16 \sum\limits_{i,j,l = 1}^{2N} L^{(0)*}_{\alpha i } R^{(0)}_{\beta j} L^{(0)*}_{\beta j} R^{(0)}_{\alpha l}
\left( D^{(\xi)}_{ij} D^{(\xi)}_{jl} + D^{(\eta)}_{ij} D^{(\eta)}_{jl} \right)
\nonumber \\
&&=  \frac16 \sum\limits_{j = 1}^{2N} R^{(0)}_{\beta j} L^{(0)*}_{\beta j}
\left[ \left( \sum_i L^{(0)*}_{\alpha i } D^{(\xi)}_{ij} \right) \left( \sum_l D^{(\xi)}_{jl} R^{(0)}_{\alpha l} \right)
\right.
\nonumber \\
&&+ \left. \left( \sum_i L^{(0)*}_{\alpha i } D^{(\eta)}_{ij} \right) \left( \sum_l D^{(\eta)}_{jl} R^{(0)}_{\alpha l} \right) \right]
\nonumber \\
&&=  \frac16 \sum\limits_{j = 1}^{2N} R^{(0)}_{\beta j} L^{(0)*}_{\beta j}
\left[ \left( \hat L^{(0)*} \hat D^{(\xi)} \right)_{\alpha j} \left( \hat D^{(\xi)} \hat R^{(0)T} \right)_{j \alpha}
\right.
\nonumber \\
&&+ \left. \left( \hat L^{(0)*} \hat D^{(\eta)} \right)_{\alpha j} \left( \hat D^{(\eta)} \hat R^{(0)T} \right)_{j \alpha} \right]
\label{term7}
\end{eqnarray}

Even terms in Eq.\ (\ref{lambda2av2a}) can be transformed in a similar way and reduce to equations of the same form as Eqs.\ (\ref{term1}--\ref{term7}) with some of indices $i$ and $j$ replaced by $\tilde i$ and $\tilde j$, respectively. Combining all these equations, we obtain
\begin{eqnarray}
&&\langle \langle L^{(0)}_{\alpha}| \hat V^{(1)} |R^{(0)}_{\beta}\rangle \langle L^{(0)}_{\beta}| \hat V^{(1)} |R^{(0)}_{\alpha}\rangle \rangle
\nonumber \\
&& = \frac16 \sum\limits_{i = 1}^{2N} L^{(0)*}_{\alpha i }  L^{(0)*}_{\beta i}
\left[ \left( \hat D^{(\xi)} \hat R^{(0)T} \right)_{i \beta}
\left( \hat D^{(\xi)} \hat R^{(0)T} \right)_{i \alpha} 
\right.
\nonumber \\
&&+  \left. \left( \hat D^{(\eta)} \hat R^{(0)T} \right)_{i \beta}
\left( \hat D^{(\eta)} \hat R^{(0)T} \right)_{i \alpha} \right]
\nonumber \\
&& +\frac16 \sum\limits_{j = 1}^{2N} R^{(0)}_{\beta j} R^{(0)}_{\alpha j}
\left[ \left( \hat L^{(0)*} \hat D^{(\xi)} \right)_{\alpha j} \left( \hat L^{(0)*} \hat D^{(\xi)} \right)_{\beta j} 
\right.
\nonumber \\
&&+ \left. \left( \hat L^{(0)*} \hat D^{(\eta)} \right)_{\alpha j} \left( \hat L^{(0)*} \hat D^{(\eta)} \right)_{\beta j} \right]
\nonumber \\
&&-\frac16 \sum\limits_{i = 1}^{2N} L^{(0)*}_{\alpha i }  R^{(0)}_{\alpha i}
\left[ \left( \hat D^{(\xi)} R^{(0)T} \right)_{i \beta} \left( \hat L^{(0)*} \hat D^{(\xi)} \right)_{\beta i}
\right.
\nonumber \\
&&+ \left. \left( \hat D^{(\eta)} R^{(0)T} \right)_{i \beta} \left( \hat L^{(0)*} \hat D^{(\eta)} \right)_{\beta i} \right]
\nonumber \\
&&-\frac16 \sum\limits_{j = 1}^{2N} R^{(0)}_{\beta j} L^{(0)*}_{\beta j}
\left[ \left( \hat L^{(0)*} \hat D^{(\xi)} \right)_{\alpha j} \left( \hat D^{(\xi)} \hat R^{(0)T} \right)_{j \alpha}
\right.
\nonumber \\
&&+ \left. \left( \hat L^{(0)*} \hat D^{(\eta)} \right)_{\alpha j} \left( \hat D^{(\eta)} \hat R^{(0)T} \right)_{j \alpha} \right]
\nonumber \\
&& + \frac16 \sum\limits_{i = 1}^{2N} L^{(0)*}_{\alpha i }  L^{(0)*}_{\beta \tilde i}
\left[ \left( \hat D^{(\xi)} \hat R^{(0)T} \right)_{i \beta}
\left( \hat D^{(\xi)} \hat R^{(0)T} \right)_{\tilde i \alpha}
\right.
\nonumber \\
&&+ \left. \left( \hat D^{(\eta)} \hat R^{(0)T} \right)_{i \beta}
\left( \hat D^{(\eta)} \hat R^{(0)T} \right)_{\tilde i \alpha} \right]
\nonumber \\
&& +\frac16 \sum\limits_{j = 1}^{2N} R^{(0)}_{\beta j} R^{(0)}_{\alpha \tilde j}
\left[ \left( \hat L^{(0)*} \hat D^{(\xi)} \right)_{\alpha j} \left( \hat L^{(0)*} \hat D^{(\xi)} \right)_{\beta \tilde j}
\right.
\nonumber \\
&&+ \left. \left( \hat L^{(0)*} \hat D^{(\eta)} \right)_{\alpha j} \left( \hat L^{(0)*} \hat D^{(\eta)} \right)_{\beta \tilde j} \right]
\nonumber \\
&&-\frac16 \sum\limits_{i = 1}^{2N} L^{(0)*}_{\alpha i }  R^{(0)}_{\alpha \tilde i}
\left[ \left( \hat D^{(\xi)} R^{(0)T} \right)_{i \beta} \left( \hat L^{(0)*} \hat D^{(\xi)} \right)_{\beta \tilde i}
\right.
\nonumber \\
&&+ \left. \left( \hat D^{(\eta)} R^{(0)T} \right)_{i \beta} \left( \hat L^{(0)*} \hat D^{(\eta)} \right)_{\beta \tilde i} \right]
\nonumber \\
&&-\frac16 \sum\limits_{j = 1}^{2N} R^{(0)}_{\beta j} L^{(0)*}_{\beta \tilde j}
\left[ \left( \hat L^{(0)*} \hat D^{(\xi)} \right)_{\alpha j} \left( \hat D^{(\xi)} \hat R^{(0)T} \right)_{\tilde j \alpha}
\right.
\nonumber \\
&&+ \left. \left( \hat L^{(0)*} \hat D^{(\eta)} \right)_{\alpha j} \left( \hat D^{(\eta)} \hat R^{(0)T} \right)_{\tilde j \alpha} \right]
\label{lambda2full}
\end{eqnarray}
Equations (\ref{lambda2av1a}) and (\ref{lambda2full}) substituted into Eqs.\ (\ref{perturb}) and (\ref{lambda2av}) yield dashed lines in Fig.\ \ref{fig50}(b). 

\bibliographystyle{apsrev4-2}
\bibliography{refstopo}

\end{document}